\def\year{2019}\relax
\documentclass[letterpaper]{article} %DO NOT CHANGE THIS
\usepackage{aaai19}  %Required
\usepackage{times}  %Required
\usepackage{helvet}  %Required
\usepackage{courier}  %Required
\usepackage{url}  %Required
\usepackage{graphicx}  %Required

\usepackage{booktabs}
\usepackage{multirow}
\usepackage{color}
\usepackage{amsmath}
\usepackage{subfigure}

\usepackage{aaai19}  %Required
\usepackage{times}  %Required
\usepackage{helvet}  %Required
\usepackage{courier}  %Required
\usepackage{amsmath}
\usepackage{amssymb}
\usepackage{url}  %Required
 \usepackage{comment}
\usepackage{graphicx}  %Required
\begin{document}
\title{AAG-Stega: Automatic Audio Generation-based Steganography}
\author{Zhongliang Yang, Xingjian Du, Yilin Tan, Yongfeng Huang, Yu-Jin Zhang\\
Department of Electronic Engineering, Tsinghua University, Beijing, 100084, China.\\
International School, Beijing University of Posts and Telecommunications, Beijing, 100876, China.\\
School of Computer Engineering and Science, Shanghai University, Shanghai, 200444, China.\\
E-mail: yangzl15@mails.tsinghua.edu.cn, yfhuang@tsinghua.edu.cn\\
}

%\author{Paper ID: 4846
%}
\maketitle
\begin{abstract}
Steganography, as one of the three basic information security systems, has long played an important role in safeguarding the privacy and confidentiality of data in cyberspace. Audio is one of the most common means of information transmission in our daily life. Thus it's of great practical significance to use audio as a carrier of information hiding. At present, almost all audio-based information hiding methods are based on carrier modification mode. However, this mode is equivalent to adding noise to the original signal, resulting in a difference in the statistical feature distribution of the carrier before and after steganography, which impairs the concealment of the entire system. In this paper, we propose an automatic audio generation-based steganography(AAG-Stega), which can automatically generate high-quality audio covers on the basis of the secret bits stream that needs to be embedded. In the automatic audio generation process, we reasonably encode the conditional probability distribution space of each sampling point and select the corresponding signal output according to the bitstream to realize the secret information embedding. We designed several experiments to test the proposed model from the perspectives of information imperceptibility and information hidden capacity. The experimental results show that the proposed model can guarantee high hidden capacity and concealment at the same time.

\end{abstract}

\section{Introduction}

\noindent In the monograph on information security\cite{shannon1949communication}, Shannon summarized three basic information security systems: encryption system, privacy system, and concealment system. Encryption system encodes the information in a special way so that only authorized parties can decode it while those who are not authorized cannot. It ensures the security of information by making the message indecipherable. Privacy system is mainly to restrict access to information, so that only authorized users can access important information. Unauthorized users cannot access it by any means under any circumstances. However, while these two systems ensure information security, they also expose the existence and importance of information, making it more vulnerable to get attacks, such as interception and cracking\cite{Bernaille2007Early}. Concealment system is very different from these two secrecy systems. It uses various carriers to embed secret information and then transmit through public channels, hide the existence of secret information to achieve the purpose of not being easily suspected and attacked\cite{Simmons1984The}. Due to its extremely powerful information hiding ability, information hiding system plays an important role in protecting trade secrets, military security and even national defense security. 
\begin{figure}[!t]
\centering
\includegraphics[width=\linewidth,height=6cm]{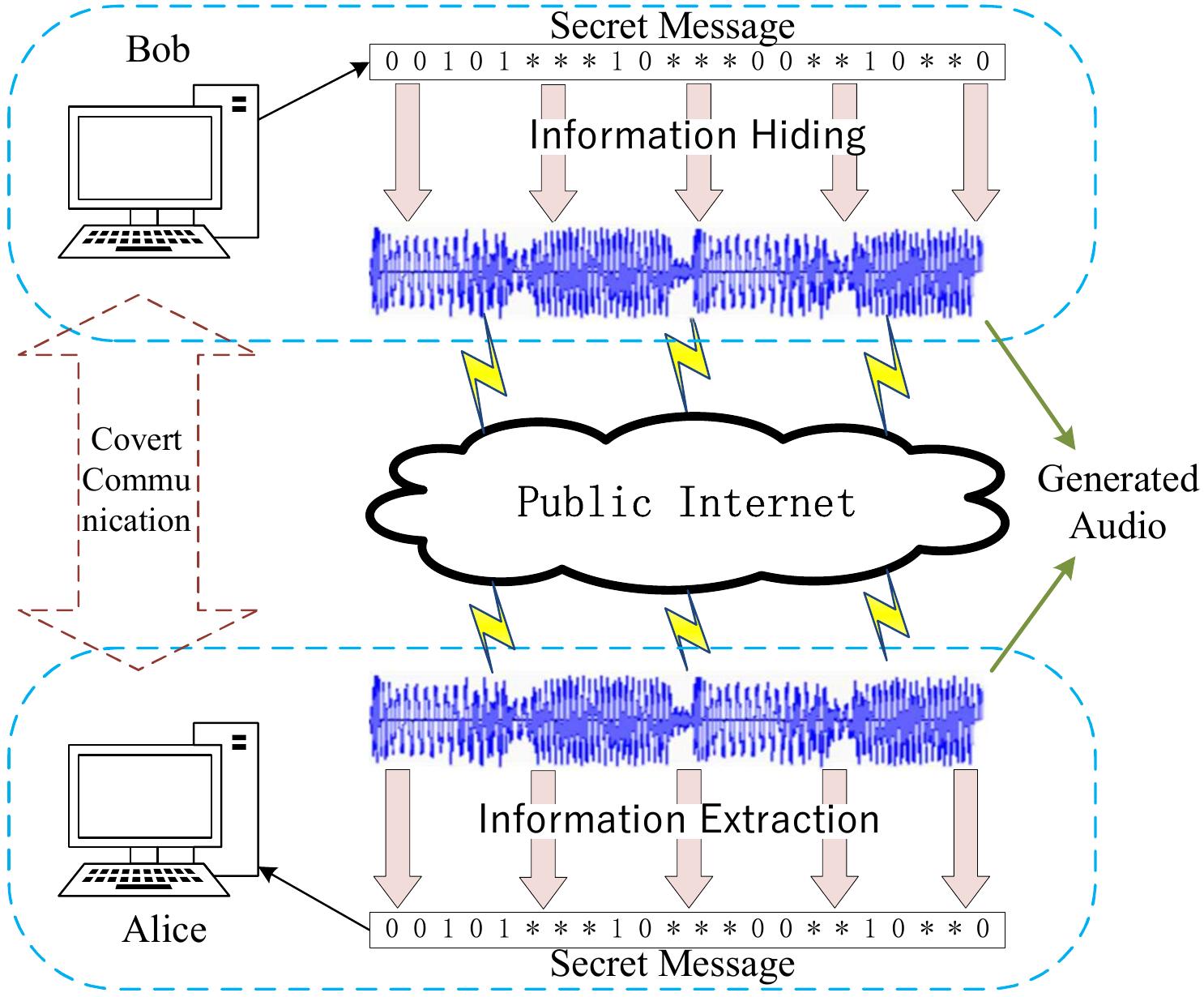}
\caption{The overall framework of the proposed method. The sender uses the proposed model to automatically generate a piece of music based on the secret bits stream that needs to be transmitted, and then sends them over the open network channel. The receiver uses the decoding algorithm to decode the received music and obtain the secret information.}
\label{fig:1}
\end{figure}

Steganography is the key technology in a concealment system. A concealment system can be illustrated by Simmons' ``Prisoners’ Problem"\cite{Simmons1984The}: Alice and Bob are in jail, locked up in separate cells far apart from each other, and they wish to devise an escape plan but cannot be perceived by the warden Eve. Faced with this situation, Alice and Bob intend to hide their true information in the normal carrier. We can model this task in a mathematical way as follows. Alice and Bob need to transmit some secret message $m$ in the secret message space $\mathcal{M}$. Alice gets a cover $x$ from the cover space $\mathcal{C}$. Under the guidance of a certain key $k_A$ in the keys space $K$, the mapping function $f$ is used to map $x$ to $s$ which is in the hidden space $\mathcal{S}$, that is:

\begin{equation}
Emb: \mathcal{C} \times \mathcal{K} \times \mathcal{M} \to \mathcal{S}, f(x,k_A,m) = s.
\end{equation}

\noindent Bob uses the extraction function $g$ to extract the correct secret message $m$ from the hidden object $s$ under the guidance of the key $k_B$ in the keys space $K$:

\begin{equation}
Ext: \mathcal{S} \times \mathcal{K} \to \mathcal{M}, g(s,k_B) = m.
\end{equation}

\noindent In order not to expose the existence of the embedded information, it is usually required that the elements in $\mathcal{S}$ and $\mathcal{C}$ are exactly the same, that is $\mathcal{S} = \mathcal{C}$. But generally speaking, this mapping function will affect the probability distributions, named $P_{\mathcal{C}}$ and $P_{\mathcal{S}}$. In order to prevent suspicion, we generally hope that the steganographic operation will not cause big differences in the probability distribution space of the carrier, that is:

\begin{equation}
d_f(P_{\mathcal{C}},P_{\mathcal{S}}) \leq \varepsilon.
\end{equation}

There are various media forms of carrier that can be used for information hiding, including image\cite{fridrich2009steganography}, audio\cite{yang2017sudoku,huang2011steganography}, text\cite{Luo2017Text,fang2017generating} and so on\cite{johnson2008detection}. Audio is a main carrier of human communication and information transmission, but it is also very easy to be monitored. Therefore, it is of great significance to study audio steganography and find an effective way to use audio carriers transmit secret messages and ensure information security. However, audio steganography is considered more difficult than image steganography because the Human Auditory System (HAS) is more sensitive than Human Visual System (HVS)\cite{gopalan2003audio}. For the above reasons, audio steganography has attracted a large number of researchers' interests. In recent years, more and more audio based information hiding methods have emerged\cite{yang2017sudoku,gupta2014dwt,huang2011steganography}. 

Fridrich J\cite{fridrich2009steganography} has summarized that, in general, steganography algorithms can utilize three different fundamental architectures that determine the internal mechanism of the embedding and extraction algorithms: steganography by cover selection, cover modification, and cover synthesis. In steganography by cover selection, Alice first encodes all the covers in a cover set and then selects different covers for transmission to achieve the covert message delivery. The advantage of this approach is that the cover is always ``100\% natural", but an obvious disadvantage is an impractically low payload. The most studied steganography paradigm today is the steganography by cover modification. Alice implements the embedding of secret information by modifying a given carrier. This kind of method has a wide range of applications on multiple carriers such as images\cite{fridrich2009steganography}, speeches\cite{huang2011steganography}, and texts\cite{topkara2006hiding}. However, directly modifying the carrier usually affects the statistical distribution of the carrier, making it easy to be detected by Eve. The third type of method is steganography by cover synthesis. Alice automatically generates a carrier based on the secret message that needs to be delivered, and embeds covert information during the generation process. The biggest advantage of this method is that it does not need to be given a carrier in advance, but can directly generate a carrier that conforms to the corresponding statistical distribution. Therefore, it has a broader application prospect than the first two methods, and is considered a very promising research direction in the current steganography field. There has been methods based on automatic generation of vectors in the field of text steganography\cite{Luo2017Text,fang2017generating} and image steganography\cite{hayes2017generating}. However, to the best of our knowledge, we have not found an effective information hiding method based on automatic audio generation. 

%Perhaps the method proposed in \cite{sampat2012audio} is the only attempt to do so, but the scheme they proposed relied too much on artificially set rules, and the model was simple and impractical.

In this paper, we propose an audio steganography based on Recurrent Neural Networks (AAG-Stega), which belongs to the third category that can automatically generate high-quality audios based on secret bits stream that needs to be embedded. In the audio generation process, we code each note reasonably based on their conditional probability distribution, and then control the audio generation according to the bits stream. We can finely adjust the encoding part to control the information embedding rate, so that we can ensure that the concealment and hidden capacity can be optimized at the same time through fine control. In this way, our model can guarantee a good enough concealment while achieving a high hidden capacity.

\section{Related Work}

\subsection{Audio Steganography}

Most of the previous audio steganography is base on the carrier modification mode, the difference is the modified features and ways. Currently, the most commonly used audio steganography methods include Least Significant Bit(LSB) encoding, phase coding, echo hiding, and Spread Spectrum(SS) method\cite{jayaram2011information}.

The basic idea of the LSB encoding is to replace the least significant bit of the cover file to hide a sequence of bytes containing the hidden data\cite{chowdhury2016view}. This type of method is simple and easy to implement. However, at the same time its fatal shortcoming is poor anti-attack ability, such as channel interference, data compression, filtering, etc. will destroy hidden information. Phase encoding\cite{bender1996techniques} mainly uses the human ear's insensitivity to absolute phase, replaces the absolute phase of the original audio segment with the reference phase representing the secret information, and adjusts the other audio segments to maintain the relative phase between the segments. This method has little effect on the original audio and is hard to detect. However, the hidden capacity is small, and when the reference phase indicating the secret information changes abruptly, a significant phase difference occurs. The basic principle of echo hiding is to embed secret information into the original audio by introducing echoes\cite{ghasemzadeh2015toward}. It takes advantage of the fact that the human auditory system cannot detect short echoes (milliseconds). The echo hiding method has strong robustness and the ability to resist active attacks. In the case of reasonable selection of echo parameters, the additional echo is difficult to be perceived by the human auditory system. For spread spectrum (SS) method\cite{kaur2015enhanced}, it spreads the secret information to the widest possible spectrum or the specified frequency band by using a spread spectrum sequence that is independent of the hidden information to achieve the purpose of information hiding. However, the SS method shares a disadvantage with LSB and parity coding in that it can bring noise into the sound file.

These above methods require a given audio carrier in advance, and then achieve information hiding by modifying some of the relatively insensitive features. However, these modifications are equivalent to adding noise to the original signal, resulting in a large difference in the statistical distribution of the carrier before and after information hiding. Such methods inevitably lead to an irreconcilable contradiction between concealment and hidden capacity. If they want to satisfy the formula (3), it will greatly limit the scope and extent of the modification, affecting the hidden capacity. Once the performance (such as hidden capacity) is improved, the performance (such as concealment) on the other hand will be greatly impaired.

However, the steganography method based on the automatic generation of information carriers does not need to be given a carrier in advance, but can automatically generate a piece of information carrier according to the covert information. In the process of generation, it can learn the statistical model of a large number of samples and generate a steganographic carrier that conforms to its statistical distribution, so it can effectively alleviate or avoid this dilemma, that is, achieve high concealment and high-capacity information hiding at the same time. For this reason, the steganography based on the automatic generation of information carriers is considered to be a very promising research direction in the field of steganography.

%In fact, the use of automatic media generation technology to achieve information hiding is not a new concept. For example, in the early stage, Chapman \emph{et al.}\cite{chapman1997hiding} tried to use syntactic template or syntax structure tree to generate steganographic texts, they hope the generated texts could conform these syntactic rules. However, such early work was limited by the automatic media generation technology, and the steganographic vectors they generated were of low quality and were easily identified, so they were not widely used. In recent years, with the development of deep learning technology, media automatic generation technology has achieved leap-forward development, and the computer automatically generated media has reached a very realistic level, such as images, text, audio and so on. With the innovation of these technologies, steganography based on vector auto-generation has re-emerged the attention of relevant researchers and has been able to generate enough realistic steganographic vectors. At present, steganography based on automatic generation of vectors is mainly concentrated in the field of images and texts. Audio is one of the most commonly used methods of information transmission in daily life, however, to the best of our knowledge, we have not found an information hiding method based on automatic audio generation.

\subsection{Audio automatic generation based on RNN}

\begin{figure}[ht]
\centering
\includegraphics[width=\linewidth]{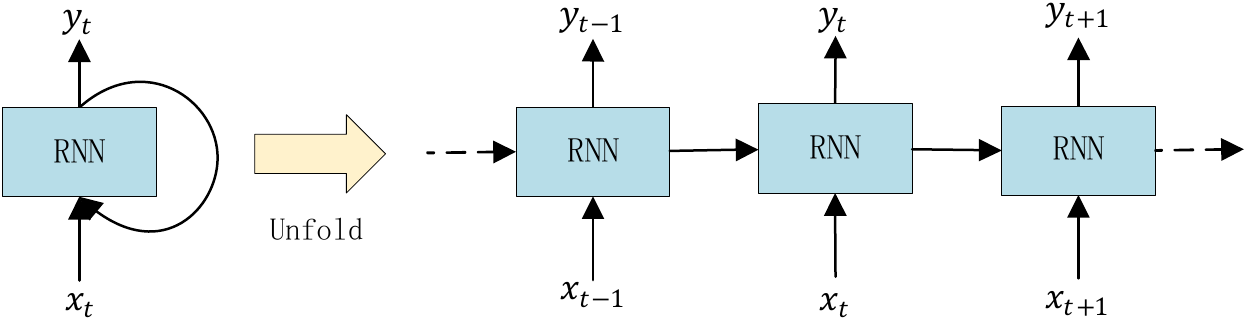}
\caption{The structure of RNN.}
\label{fig:2}
\end{figure}

Automatic generation of sufficiently realistic information carriers has always been hard. In recent years, with the development of deep neural network technology, more and more research works proved that we can use the powerful feature extraction and expression capabilities of deep neural networks to model information carriers and generate real enough covers such as image, text and audio. Recurrent Neural Network\cite{mikolov2010recurrent} is a special artificial neural network model. Unlike other deep neural networks, RNN is actually not ``deep" in space, the simplest RNN can have only one hidden layer. The fundamental feature of a Recurrent Neural Network is that the network contains a feed-back connection at each step, so it can be extended in the time dimension and form a ``deep" neural network in time dimension, which has been shown in Figure \ref{fig:2}. 

Due to its special structural characteristics, RNN is very suitable for modeling sequential signals. Audio is a typical sequence of signals. For a piece of audio, we can represent it as a sequence $S=\{x_1,x_2,...,x_n\}$, with each element in the sequence $S$ representing the signal at each moment. Currently, the majority of audio automatic generation work is modeled in such a way.  The signal of each time $t$ can be expressed as a conditional probability distribution based on the first $t-1$ time signals, the probability distribution of the entire audio can be expressed as the product of the $n$ conditional probabilities, which can be expressed by the following formulas:

\begin{equation}
\begin{aligned}
& p(S) = p(x_1,x_2,x_3,...,x_n)\\
& = p(x_1)p(x_2\mid x_1)...p(x_n\mid x_1,x_2,...,x_{n-1}),
\end{aligned}
\end{equation}

With the powerful self-learning ability of neural networks, when we provide enough training samples, the model will have the ability to automatically learn and get the optimal conditional probability distribution estimate. In this way, we can iteratively select the signal with the highest conditional probability as the output, and finally generate a sufficiently realistic information carrier. A lot of preliminary work has proven that it is effective to use this method for automatic audio generation\cite{sturm2016music}.

\begin{figure*}[ht]
\centering
\includegraphics[width=\linewidth,height=8.5cm]{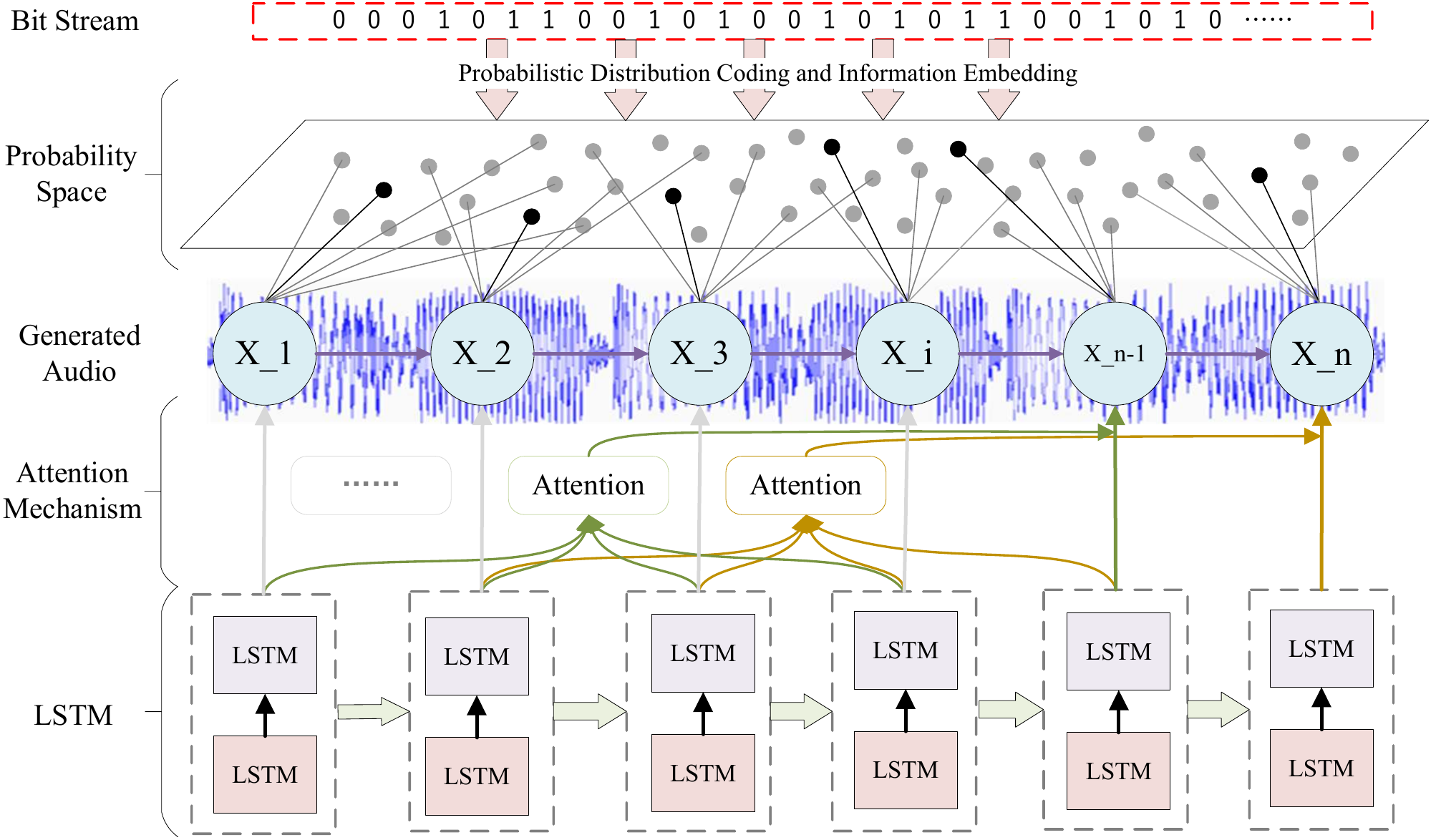}
\caption{A detailed explanation of the proposed model and the information hiding algorithm. The top of the figure is the bits stream that needs to be embedded. The middle part is the generated steganographic audio, and the bottom is a two-layer LSTM with the Lookback mechanism and the Attention mechanism. We encode the probability distribution of notes, and then select the corresponding note according to the secret bitstream, so as to achieve the purpose of hiding information.}
\label{fig:3}
\end{figure*}

\section{AAG-Stega Methodology}

Compared to other steganographic modes, the steganography methods based on carrier automatic generation are characterized by the fact that they do not need to be given carrier in advance. Instead, they can automatically generate a textual carrier based on secret information. This type of method omits $\mathcal{C}$ in Equation (1), and the embedding function becomes: 

\begin{equation}
Emb: \mathcal{K} \times \mathcal{M} \to \mathcal{S}, f(k_A,m) = s.
\end{equation}

\noindent However, they still have to satisfy formula (3), that is, the steganographic operation should minimize the impact of the carrier on the semantic spatial distribution. This could be very hard, but also very promising. The overall structure of our model is shown in Figure \ref{fig:3}. The whole system consists of three modules: automatic audio generation (AAG) module, information embedding module and information extraction module.

\subsection{AAG Module}

By modeling the audio as a product of the conditional probabilities at each moments, as shown in equation (4), it is easy to know that if we want to generate high-quality audio, we need to get an optimal estimate of the conditional probability distribution at each moment, than is $p(x_n\mid x_1,x_2,...,x_{n-1})$.

For the simplest recurrent neural network that has only one hidden layer, it can be described in the following set of formulas:

\begin{equation}
\left\{\begin{array}{l}

h_t = f_h(W_h \cdot x_t + U_t \cdot h_{t-1} + b_h), \\
y_t = f_o(W_o \cdot h_t + b_o),\\
\end{array}  
        \right.
\end{equation}

\noindent where $x_t$ and $y_t$ indicate the input and output vector at $t$-th step respectively, $h_t$ represents the vector of hidden layer, $W_.$, $U_.$ and $b.$ are learned weight matrices and biases, $f_h$ and $f_o$ are nonlinear functions, where we usually use $tanh$ or $softmax$ function.

Theoretically, this simplest RNN model can deal with arbitrary length sequence signals. However, due to the gradient vanish problem\cite{Hochreiter1998The}, it cannot handle with the problem of long-range dependence effectively. But its improved algorithm, Long Short-Term Memory (LSTM) model\cite{Hochreiter1997Long}, can effectively solve this problem by elaborately designed unit nodes. The main improvement of LSTM is the hidden layer unit, which is composed of four components: a cell, an input gate, an output gate and a forget gate. It can store the input information of the past time into the cell unit so as to overcome the problem of long distance dependence, and realize the modeling of long time series. An LSTM unit can be described using the following formulas:

\begin{flalign}
& \left\{\begin{array}{l}
 I_t = \sigma(W_i \cdot [h_{t-1},x_t] + b_i), \\
 F_t = \sigma(W_f \cdot [h_{t-1},x_t] + b_f),\\
 C_t = F_t \cdot C_{t-1} + I_t \cdot \tanh(W_c [h_{t-1},x_t] + b_c)\\
 O_t = \sigma(W_o \cdot [h_{t-1},x_t] + b_o),\\
 h_t = O_t \cdot \tanh(C_t).\\
             \end{array}  
        \right.&
\end{flalign}

\noindent where $I_t$ indicates the input gate, it controls the amount of new information to be stored to the memory cell. The forget gate, which is $F_t$, enables the memory cell to throw away previously stored information. In this regard, the memory cell $C_t$ is a summation of the incoming information modulated by the input gate and previous memory modulated by the forget gate $F_t$ . The output gate $O_t$ allows the memory cell to have an effect on the current hidden state and output or block its influence.

Basic RNN or LSTM can only generate a short melody that stays in key, but they have trouble generating a longer melody. In order to improve the model's ability to learn longer-term structures, in our automatic audio generation module, we use the Lookback mechanism\cite{lookback_attention_rnn} and the Attention mechanism\cite{Bahdanau2014Neural}. For the basic RNN, the input to the model was a one-hot vector of the previous event, and the label was the target next event. In Lookback RNN, some additional information will be added to the input vector: 1) the events from 1 and 2 bars ago, 2) signals if the last event was creating something new, or just repeating an already established melody, 3) the current position within the measure.

To learn even longer-term structure, we further combine the attention mechanism. Attention is one of the ways that models can access previous information without having to store it in the RNN cell’s state. The difference is that our model does not pay attention to the output of all the previous moments, but only the outputs from the last $m$ steps when generating the output for the current step, which can be described in the following formulas:

\begin{equation}
\begin{aligned}
& e_i^t = f_{att}(h_i, c_t),\\
& \alpha_i^t = \frac{exp(e_i^t)}{\sum^{t-1}_{k=t-m}{exp(e_k^t)}},\\
& z_t = \sum^{t-1}_{k = t-n}\alpha_k^t \cdot h_k,
\end{aligned}
\end{equation}

\noindent where $f_{att}$ is a multilayer perceptron conditioned on the hidden state of last $n$ steps. $h_i$ are the RNN outputs from the previous $n$ steps, and vector $c_t$ is the current step’s RNN cell state. These values are used to calculate $e_i^t$, an $m$ length vector with one value for each of the previous $m$ steps. The values represent how much attention each step should receive. A softmax is used to normalize these values and create a mask-like vector $\alpha_i^t$, called the attention mask. The RNN outputs from the previous $m$ steps are then multiplied by these attention mask values and then summed together to get $z_t$. 

The $z_t$ vector is essentially all $m$ previous outputs combined together, but each output contributing a different amount relative to how much attention that step received. This $z_t$ vector is then concatenated with the RNN output from the current step and a linear layer is applied to that concatenated vector to create the new output for the current step. Different from some other attention models that only apply this $z_t$ vector to the RNN output, in our module, this $z_t$ vector is also applied to the input of the next step. The $z_t$ vector is concatenated with the next step’s input vector and a linear layer is applied to that concatenated vector to create the new input to the RNN cell. This helps attention not only affect the data coming out of the RNN cell, but also the data being fed into the RNN cell.

As we have mentioned before, the output at time step $t$ is not only based on the input vector of the current time $x_t$, but also the information of the previous $(t-m)$ moments. Therefore, the output of the last hidden layer at time step $t$ can be regarded as the information fusion of the previous $t-m$ steps. Based on these features, after all the hidden layers, we add a softmax layer to calculate the probability distribution of the $t$-th value, that is

\begin{equation}
p(x_{t+1}) \propto \exp(L_0(Ex_{t-1}+L_hh_t+L_zz_t)),
\end{equation}

\noindent where $L_0$, $E$, $L_h$, $L_z$ are learned parameters. All the parameters of the neural network need to be obtained through training. In the training process, we update network parameters using softmax cross-entropy loss and backpropagation algorithm\cite{Rumelhart1988Learning}. After minimizing the loss function through the iterative optimization of the network, we will get a good estimate of the probability distribution in Equation (4).

\subsection{Information Hiding Algorithm}

In the information embedding module, we mainly code each note based on their conditional probability distribution, which is $p(x_n\mid x_1,x_2,...,x_{n-1})$, to form the mapping relationship from the binary bitstream to note space. Our thought is mainly based on the fact that when our model is well trained, there is actually more than one feasible solution at each time point. After descending the prediction probability of all the notes in the dictionary $D$, we can choose the top $m$ sorted notes to build the Candidate Pool (CP). To be more specific, suppose we use $c_i$ to represent the $i$-th note in the Candidate Pool, then the CP can be written as

$$CP = [c_1,c_2,...,c_m].$$

In fact, when we choose a suitable size of the Candidate Pool, any note $c_i$ in CP selected as the output at that time step is reasonable and will not affect the quality of the generated audio, so it becomes a place where information can be embedded. It is worth noting that at each moment we choose different note, the probability distribution of next note will be different according to the Equation$(4)$. After we get the Candidate Pool, we need to find an effective encoding method to encode the words in it.

\begin{figure}[ht]
\centering
\includegraphics[width=\linewidth]{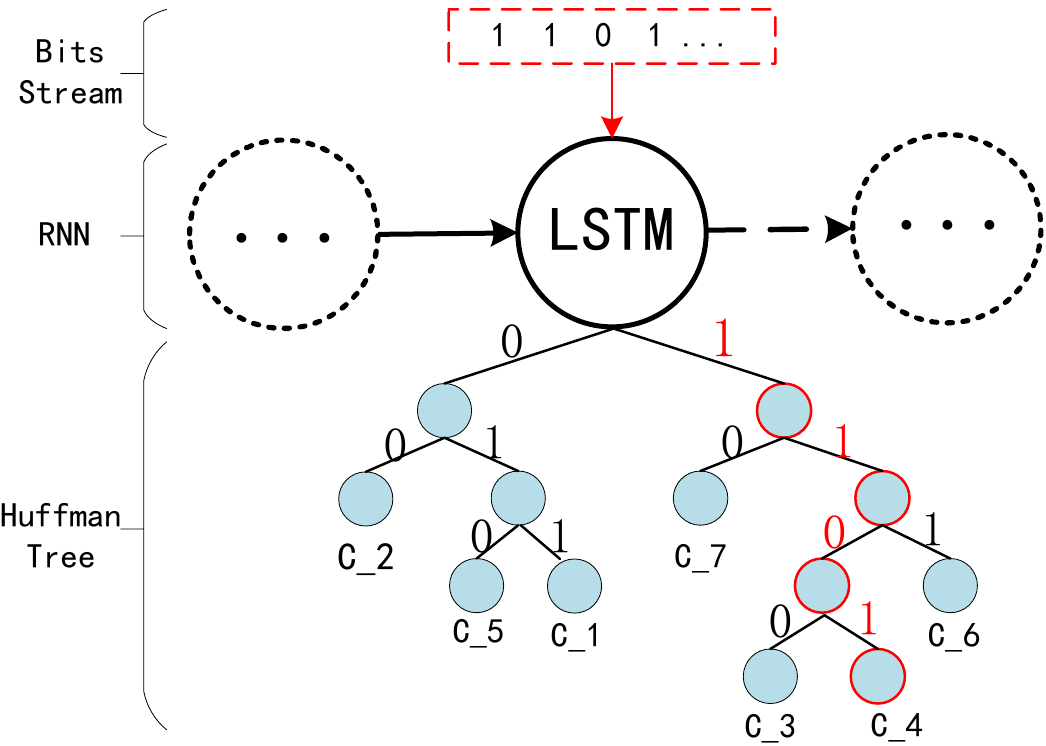}
\caption{Encoding notes in Candidate Pool using a Huffman tree.}
\label{fig:4}
\end{figure}

In order to make the coding of each word more in line with its conditional probability distribution, we use the Huffman tree to encode the words in the candidate pool. In computer science and information theory, a Huffman code is a particular type of optimal prefix code. The output from Huffman's algorithm can be viewed as a variable-length code table for encoding a source symbol. In the encoding process, this method takes more into consideration of the probability distribution of each source symbol in the construction process, and can ensure that the code length required by the symbol with higher coding probability is shorter\cite{huffman1952method}. In the audio generation process, at each moment, we represent each note in the Candidate Pool with each leaf node of the tree, the edges connect each non-leaf node (including the root node) and its two child nodes are then encoded with 0 and 1, respectively, with 0 on the left and 1 on the right, which has been shown in Figure 4. 

After the notes in the Candidate Pool are all encoded, the process of information embedding is to select the corresponding leaf node as the output of the current time according to the binary code stream that needs to be embedded. Algorithm details of the proposed information hiding method are shown in Algorithm 1. With this method, we can generate a piece of natural audio according to the input secret code stream. And then these generated audio can be sent out through the open channel to achieve the purpose of secret information hidden and sent, which has a high concealment.

{\tabcolsep=4pt\small
\begin{center}
\begin{tabular}{lp{82mm}} \hline
 \multicolumn{2}{l}{{\bf Algorithm 1}~~ Information Hiding Algorithm}\\ \hline
    &\hspace*{0.02in} {\bf Input:}\\
    &\hspace*{0.1in}Secret bit stream: $B = \{0,0,1,0,1,...,0,1,0\}$\\
    &\hspace*{0.1in}Size of Candidate Pool(CPS): $m$\\
    &\hspace*{0.1in}Start notes list: $A = \{key_1,key_2,...,key_F\}$\\
    &\hspace*{0.02in} {\bf Output:}\\
    &\hspace*{0.1in}Generated Steganographic Audio: $S = \{x_1,x_2,...,x_N\}$\\
    &\quad if (not the end of current audio) then\\
    &\qquad Calculate the probability distribution of the next note according to the previously generated notes using well trained RNN;\\
    &\qquad Descending the prediction probability of all the notes and select the top $m$ sorted notes to form the Candidate Pool(CP);\\
    &\qquad Construct a Huffman tree according to the probability distribution of each notes in the CP and encode the tree;\\
    &\qquad Read the binary stream, and search from the root of the tree according to the encoding rules until the corresponding leaf node is found and output its corresponding note;\\
    &\quad else\\
    &\qquad Random select a start note $key_i$ in the Start notes list $A$ as the start of the next audio;\\
    &\quad end if \\
    &\hspace*{0.1in} {\bf Return:} Generated Steganographic Audio\\
 \hline
\end{tabular}
\end{center}}

\subsection{Information Extraction Algorithm}

Information embedding and extraction are two completely opposite operations. After receiving the steganographic audio, the receiver inputs the first note of each audio as a key into the RNN which will calculate the distribution probability of the notes at each subsequent time point in turn. At each time point, when the receiver gets the probability distribution of the current note, he firstly sorts all the notes in the dictionary in descending order of probability, and selects the top $m$ notes to form the Candidate Pool. Then he builds Huffman tree according with the same rules to encode the notes in the candidate pool. Finally, according to the actual transmitted word at the current moment, the path of the corresponding leaf node to the root node is determined, so that we can successfully and accurately decode the bits embedded in the current note. By this way, the bits stream embedded in the original audio can be extracted very quickly and without errors.
%Algorithm details of the proposed information extraction method are shown as Algorithm 2.

\begin{comment}
{\tabcolsep=4pt\small
\begin{center}
\begin{tabular}{lp{82mm}} \hline
 \multicolumn{2}{l}{{\bf Algorithm 2}~~ Information Extraction Algorithm}\\ \hline
    &\hspace*{0.02in} {\bf Input:}\\
    &\hspace*{0.1in}Received Steganographic Audio: $S = \{x_1,x_2,...,x_N\}$.\\
    &\hspace*{0.1in}Size of Candidate Pool(CPS): $m$\\
    &\hspace*{0.02in} {\bf Output:}\\
    &\hspace*{0.1in}Secret bits stream: $B = \{0,0,1,0,1,...,0,1,0\}$.\\
    &\quad if (not the end of current audio) then\\
    &\qquad Calculate the probability distribution of the next note according to the previously notes using RNN;\\
    &\qquad Descending the prediction probability of all the notes and select the top $m$ sorted notes to form the Candidate Pool(CP);\\
    &\qquad Construct a Huffman tree according to the probability distribution of each note in the CP and encode the tree;\\
    &\qquad Determine the path from the root node to the leaf node which corresponding to the note at the current moment;\\
    &\qquad According to the tree coding rule, ie, the left side of the child node is 0 and the right side is 1, the code stream embedded in the current note is decoded;\\
    &\qquad Output the decoded code and append it to $B$;\\
    &\quad end if \\
    &\quad Read next audio; \\
    &\hspace*{0.1in} {\bf Return:} Extracted secret bits stream B\\
 \hline
\end{tabular}
\end{center}}

\end{comment}

\section{Experiments and Analysis}

In this section, we first introduce the training dataset as well as model details. Then we tested and analyzed the performance of the proposed model from two aspects: information imperceptibility and hiding capacity.

\subsection{Data Preparing and Model Detail}

For model training, we use Lakh MIDI dataset\cite{raffel2016learning} as our training dataset, it contains 176,581 unique MIDI files, 45,129 of which have been matched and aligned to entries in the Million Song Dataset\cite{bertin2011million}. 

Almost all the parameters in our model can be obtained through training, but there are still some hyper-parameters need to be determined. Through the comparison test, the hyper-parameters of our model are set as follows: the number of LSTM hidden layers is $2$, with each layer containing $256$ LSTM units. The length of attention vector is $40$ in this experimental. During model training, in order to strengthen the regularization and prevent overfitting, we adopt the dropout mechanism\cite{Srivastava2014Dropout} during the training process. We chose Adam\cite{Kingma2014Adam} as the optimization method. The learning rates are initially set as 0.001, batch size is set as 64, dropout rate is 0.5 and the attention length is 40. 

\subsection{Information Hiding Capacity}

Embedding Rate(ER) calculates how much information can be embedded in the audio. The calculation method of embedding rate is to divide the actual number of embedded bits by the number of bits occupied by the entire generated audio in the computer. The mathematical expression is as follows:

\begin{equation}
\begin{aligned}
ER &= \frac{1}{N}\sum^N_{i=1}\frac{(L_i-1)\cdot k}{B{(s_i)}} = \frac{(\overline{L}-1)\times k}{B{(s_i)}},
\end{aligned}
\end{equation}

\noindent where $N$ is the number of generated audios and $L_i$ is the number of notes in the $i$-th audio. $k$ indicates the number of bits embedded in each note and $B(s_i)$ indicates the number of bits occupied by the $i$-th audio in the computer. Obviously, the size of the candidate pool (CPS) can directly affect the embedding rate. In the experiment, for each CPS, we generated 50 pieces of audio, and counted the average number of notes contained in it and the average bytes of the generated audio. The final statistical results and the calculated information embedding rate are shown in Table 1.

\begin{table}[h]
\centering
\begin{tabular}{l|c|c|c|c|c|c}
\toprule[1.5pt]
CPS &2 &4 &8 &16 &32 &64\\
\hline
$\overline{k}$ &1 &1.95 &2.78 &3.59 &4.37 &5.22\\
\hline
$\overline{L}$ &147.9 &146.3 &146.9 &160.5 &139.5 &141.8\\
\hline
$\overline{B}(s)$ &505.8 &518.2 &524.9 &504.5 &530.8 &495.4\\
\hline
ER  &3.7\% &6.9\% &9.7\% &14.3\% &14.4\% &18.7\%\\
\bottomrule[1.5pt] 
\end{tabular}
\caption{\label{tab:1}The calculated information embedding rate under different CPS.}
\end{table}

\subsection{Information Imperceptibility}

The purpose of a concealment system is to hide the existence of information in the carrier to ensure the security of important information. Therefore, the imperceptibility of information is the most important performance evaluation factor of a concealment system. Currently, there are two different evaluation methods to measure the quality of speech, i.e., objective evaluation and subjective evaluation respectively. 

\subsubsection{Objective Evaluation}

In order to ensure a sufficiently high concealment, according to formula (3), the statistical distribution of the carriers before and after steganography is required to be as consistent as possible. For each of the 50 piece of audio generated under each CPS, we calculated their average likelihood probability as follows:

\begin{equation}
\begin{aligned}
&score = -\frac{1}{N}\sum^N_{i=1} \frac{1}{L_i-1} log{p(s_i)}\\
&= -\frac{1}{N}\sum^N_{i=1} \frac{1}{L_i-1}log(\prod^{L_i}_{j=1}p(x_j\mid x_1,x_2,...,x_{j-1})),
\end{aligned}
\end{equation}

\noindent where $N$ is the number of samples and $L_i$ indicates the number of notes in $i$-th audio. The test results are shown in table 2.

\begin{table}[h]
\renewcommand\arraystretch{1.3}
\centering
\begin{tabular}{l|c|c|c|c|c|c}
\toprule[1.5pt]
CPS &2 &4 &8 &16 &32 &64\\
\hline
score  &0.306 &0.417 &0.315 &0.503 &0.547 &0.783\\
\bottomrule[1.5pt] 
\end{tabular}
\caption{\label{tab:2}Results of the average likelihood probability.}
\end{table}

From Table 2, we can find that the calculated values are relatively small, indicating that the steganographic samples and training samples generated by our model are close in probability distribution. Although the score will gradually increase as the embedding rate increases, it remains within a reasonable range.

\subsubsection{Subjective Evaluation}

The ``A/B/X" testing is a standard subjective evaluation method in the field of audio steganography\cite{huang2011steganography}. The detail of this testing is as follows. Suppose there are three types of speech files, denoted by A, B, and X, respectively. A represents the stego audio file containing hidden information, B denotes the audio file without any hidden information, and X is either A or B. Evaluators were employed to listen the audio files, and then asked to decide whether X is A or B.

\begin{table}[h]
\renewcommand\arraystretch{1.3}
\centering
\begin{tabular}{c|c|c|c}
\toprule[1.5pt]
Evaluator &Accuracy & Recall & f1-score\\
\hline
Group1 &0.4861$\pm$0.049 &0.480$\pm$0.082 &0.585$\pm$0.064\\
Group2 &0.388$\pm$0.085 &0.389$\pm$0.161 &0.480$\pm$0.129\\
Group3 &0.475$\pm$0.063 &0.486$\pm$0.079 &0.584$\pm$0.067\\
Group4 &0.432$\pm$0.100 &0.406$\pm$0.143 &0.514$\pm$0.126\\
Group5 &0.440$\pm$0.055 &0.444$\pm$0.064 &0.548$\pm$0.057\\
\hline
Total &0.440$\pm$0.077 &0.437$\pm$0.107 &0.539$\pm$0.094\\
\bottomrule[1.5pt] 
\end{tabular}
\caption{\label{tab:3}Testing results of five evaluator groups.}
\end{table}

We invited 5 different test groups, each group contains 10 people. For each group, we randomly selected 10 steganographic audio generated by the proposed model under different CPS. At the same time, we random selected 15 audio samples without hidden information. We mixed them up to form a test dataset containing 65 audio samples. Each evaluator listens to the 65 audios one by one and gives a judgment on each audio, whether it belongs to A or belongs to B. To help them judge, we gave them 3 audio without covert information and 3 audio generated under $CPS=8$. This test data set can be found in the supplemental materials\footnote{\url{https://github.com/YangzlTHU/AAG-Stega}}.

We first calculated the accuracy, recall, and F1 score for steganographic and non-steganographic audio to be correctly identified. We uniformly marked the steganographic audio at different embedding rates as negative samples, and the audio without steganographic information as positive samples. The final test results are shown in Table 3. From Table 3, we can find that the accuracy, recall rate or F1 score are all around 0.5. This means it was impossible to distinguish the stego speech from the original speech by using the A/B/X testing method when secret information was embedded in audio using the proposed method. 

\begin{table}[h]
\centering
\begin{tabular}{c|c|c|c|c|c}
\toprule[1.5pt]
CPS &2 &4 &8 &16 &32\\
\hline
Group1 &44\% &48\% &56\% &52\% &56\%\\
Group2 &66\% &48\% &62\% &70\% &56\%\\
Group3 &52\% &40\% &58\% &52\% &52\%\\
Group4 &70\% &62\% &60\% &54\% &50\%\\
Group5 &74\% &50\% &50\% &48\% &56\%\\
\hline
Average &61.2\% &49.6\% &57.2\% &55.2\% &54.0\%\\
\bottomrule[1.5pt] 
\end{tabular}
\caption{\label{tab:4}Percentages of failures Using A/B/X Test.}
\end{table}

Table 4 shows the percentage of failures to identify the stego audio file under different CPS. From Table 4 we can find that, first, when CPS=2, that is, when the embedding rate is the lowest, the average percentage of failure judgments is the highest, which is 61.2\%. Secondly, as the embedding rate increases, the average percentage of failure judgments does not decrease significantly, both at around 50\%. This also proves that our steganographic audios are very difficult to recognize and thus the proposed model has a high degree of concealment.

\section{Conclusion}

In this paper, we proposed an automatic audio generation-based steganography(AAG-Stega), which is completely different from the previous audio steganography. The proposed model can automatically generate high-quality audio covers on the basis of the secret bits stream that needs to be embedded. The experimental results showed that the proposed model can guarantee high hidden capacity and concealment at the same time.

%\bibliographystyle{aaai}
%\bibliography{aaai}

\end{document}